\documentclass[smallcondensed]{svjour3}    % onecolumn (standard format)
\smartqed  % flush right qed marks, e.g. at end of proof
\usepackage{graphicx}
\usepackage{fix-cm}
\begin{document}

\title{Investigating the implementation of restricted sets of multiqubit operations on distant qubits: a communication complexity perspective%\thanks{Grants or other notes
}

\titlerunning{Investigating the implementation
of restricted sets of multiqubit operations}        % if too long for running head

\author{Haozhen Situ         \and
        Daowen Qiu %etc.
}


\institute{Haozhen Situ \and Daowen Qiu \at
              Department of Computer Science, Sun Yat-sen University, Guangzhou 510006,  China \\
%              Tel.: +123-45-678910\\
 %             Fax: +123-45-678910\\
              \email{issqdw@mail.sysu.edu.cn (D. Qiu)}           %  \\
           \and
           Daowen Qiu \at
              SQIG--Instituto de Telecomunica\c{c}\~{o}es, IST, TULisbon, Av. Rovisco Pais 1049-001, Lisbon, Portugal\\
              The State Key Laboratory of Computer Science,
Institute of Software,
 Chinese  Academy of Sciences,
 Beijing 100080,  China }

\date{Received: date / Accepted: date}

\maketitle

\begin{abstract}
We propose a protocol for Alice to implement a multiqubit quantum
operation from the restricted sets on distant qubits possessed by
Bob, and then we investigate the communication complexity of the
task in different communication scenarios. By comparing with the
previous work, our protocol works without prior sharing of
entanglement, and requires less communication resources than the
previous protocol in the qubit-transmission scenario. Furthermore,
we generalize our protocol to $d$-dimensional
operations.
\end{abstract}

\section{Introduction}
\label{intro} Quantum information processing (QIP) usually involves
the implementation of quantum operations between spatially separated
qubits. For example, distributed quantum computation requires
implementing nonlocal operations on the qubits at distant nodes.
This task can always be completed via the so-called bidirectional
quantum state teleportation (BQST) \cite{HVCP01}. In BQST, all the
involved qubits are first teleported to Alice, and then Alice
applies the operation and teleports them back to other parties. Thus far, many authors \cite{HVCP01,EJPP00,CLP01,YG01,HPV02,GR05,CC05,YZG06,W06,A07,W07,ZW07,ZW08,WYZ08} have studied how to implement nonlocal
operations using prior sharing of entanglement, local operations,
and classical communication (LOCC), and these researches \cite{HVCP01,EJPP00,CLP01,YG01,HPV02,GR05,CC05,YZG06,W06,A07,W07,ZW07,ZW08,WYZ08}
have shown that some special operations on distant qubits can be implemented by LOCC
using less communication resources than they are implemented in the BQST
scheme.

Eisert {\it et al} \cite{EJPP00} investigated the minimal
communication resources that are required in the local
implementation of nonlocal quantum gates, and presented optimal
protocols for a number of important gates using prior sharing of
entanglement and LOCC. For example, 1 shared ebit and communication
of 1 cbit in each direction are both necessary and sufficient for
the nonlocal implementation of a controlled-$U$ gate; an $N$-qubit
controlled-$U$ gate with $N-1$ control qubits possessed by Bob and 1
target qubit possessed by Alice (denoted as $CU(N-1,1)$ in the
remainder of this paper) can also be implemented using 1 shared ebit
and communication of 1 cbit in each direction; an $N$-qubit
controlled-$U$ gate with $N-1$ control qubits and 1 target qubit
possessed by $N$ spatially separated parties can be implemented
using $N-1$ shared ebits and communication of $2(N-1)$ cbits.

From another point of view, the task can be completed without prior
sharing of entanglement as follows: Bob sends his qubits to Alice,
and then Alice applies the operation and sends them back to Bob. In
the case of $CU(N-1,1)$, a total of $2(N-1)$ qubits need to be
communicated (we assume that $U$ is only known to Alice). Can Alice
and Bob communicate fewer qubits to complete this task? The answer
is positive. Yang \cite{Y08} proposed a protocol for implementing
$CU(N-1,1)$ without prior sharing of entanglement. The required
communication resources are 1 qubit transmitted from Bob to Alice
and 1 cbit transmitted from Alice to Bob.

Huelga {\it et al} \cite{HPV02} showed that there are two restricted
sets of one-qubit operations that can be implemented remotely using
1 shared ebit and communication of 1 cbit in each direction. One of
these two restricted sets consists of diagonal operations and the
other one consists of antidiagonal operations:

\begin{eqnarray}
U_{diag}= \left(
\begin{array}{cc}
  e^{i\phi} & 0 \\
  0 & e^{-i\phi} \\
\end{array}
\right), U_{anti}= \left(
\begin{array}{cc}
  0 & e^{i\phi} \\
  -e^{-i\phi} & 0 \\
\end{array}
\right).\end{eqnarray}

Wang \cite{W06} generalized these restricted sets to the case of
multiqubits, and proposed a protocol for Alice to implement restrict
sets of $N$-qubit operations on Bob's qubits. Operations belonging
to the restricted sets have just one nonzero element in any column
or any row. Each restricted set is characterized by a permutation
$f$, and thus an $N$-qubit operation in the restricted set $f$ can
be expressed as:
\begin{eqnarray}
U(f,\phi)= \sum_{x=0}^{2^N-1} e^{i\phi(x)} |f(x)\rangle\langle
x|.\end{eqnarray} In this paper, we assume that Alice knows
$U(f,\phi)$ but Bob only knows which restricted set $f$ the
operation belongs to (i.e. Alice has the device to implement
$U(f,\phi)$ and Bob knows the type of Alice's device. If Bob does
not know which one of the $2^N!$ restricted sets Alice's device
belongs to, Alice should first tell Bob this information through
classical communication). The required communication resources are
$N$ shared ebits and $N$ cbits in each direction.

By hybridizing the protocol in Ref.~\cite{W06} and BQST, Zhao {\it
et al} \cite{ZW07} proposed a protocol for Alice to implement
restricted sets of $(N+M)$-qubit operations on Bob's qubits.
Operations belonging to the restricted sets are $2^N\times 2^N$
block matrices with just one nonzero block in any column or any row,
and every block is a $2^M\times 2^M$ unitary matrix. Each restricted
set is characterized by a permutation $f$, and thus an $(N+M)$-qubit
operation in the restricted set $f$ can be expressed as:
\begin{eqnarray}
U(f,G)= \sum_{x=0}^{2^N-1} |f(x)\rangle\langle x| \otimes
G(x),\end{eqnarray}where $G(x)$'s are arbitrary $2^M\times 2^M$
unitary matrices. In this paper, we assume that Alice knows $U(f,G)$
but Bob only knows which restricted set $f$ the operation belongs to
(i.e. Alice has the device to implement $U(f,G)$ and Bob knows the
type of Alice's device. If Bob does not know which one of the $2^N!$
restricted sets Alice's device belongs to, Alice should first tell
Bob this information through classical communication). To implement
$U(f,G)$ on distant $N+M$ qubits possessed by Bob, $M$ qubits need
to be teleported back and forth between the two parties, and the
required communication resources are $N+2M$ shared ebits plus $N+2M$
cbits in each direction \cite{ZW07}. If the $M$ qubits that need to
be teleported in this hybrid protocol are initially possessed by
Alice, BQST is unnecessary and the required communication resources
are $N$ shared ebits plus $N$ cbits in each direction \cite{ZW08}.
If $M=0$, this hybrid protocol is reduced to the protocol in
Ref.~\cite{W06}.

Inspired by Yang's work \cite{Y08}, we study the possibility of
implementing $U(f,G)$ on distant qubits without prior sharing of
entanglement, and consider the required communication resources in
different communication scenarios. Actually, our protocol can
complete the same tasks as those in Ref.~\cite{HPV02,W06,ZW07,ZW08}
without using prior sharing of entanglement. Our protocol requires
less communication resources than the previous protocol
\cite{HPV02,W06,ZW07,ZW08} in the qubit-transmission scenario.

The remainder of the paper is organized as follows. In
Sect.~\ref{protocol}, we propose a protocol for implementing
$U(f,G)$ on distant qubits. Sect.~\ref{analysis} contains a
communication complexity analysis of the task in different
communication scenarios. In Sect.~\ref{generalization}, we
generalize our protocol to $d$-dimensional operations. A brief
conclusion follows in Sect.~\ref{conclusion}.

\section{Our protocol}
\label{protocol} We first explain how to implement $U(f,G)$ on $N$
qubits possessed by Bob and $M$ qubits possessed by Alice, and then
we explain another two cases.

In the first case, the initial state of the $N+M$ qubits can be
written as
\begin{eqnarray}
|\varphi\rangle_{B_1\ldots B_NA_1\ldots A_M} =
\sum_{j=0}^{2^N-1}\alpha_j |j\rangle_{B_1\ldots
B_N}|\xi_j\rangle_{A_1\ldots A_M},
\end{eqnarray}where subscripts $A_1\ldots A_M$ ($B_1\ldots B_N$) denotes Alice's (Bob's) $M$ ($N$)
qubits. Our protocol consists of six steps:

(1) Bob introduces $N$ ancilla qubits $C_1\ldots C_N$ initially
prepared in the state $|0\rangle^{\otimes N}$, and then performs $N$
controlled-NOT (CNOT) gates with $B_i$ as the control qubit and
$C_i$ the target qubit. After this step, the state of the composite
system becomes
\begin{eqnarray}
\sum_{j=0}^{2^N-1}\alpha_j |j\rangle_{B_1\ldots B_N}
|j\rangle_{C_1\ldots C_N} |\xi_j\rangle_{A_1\ldots A_M}.
\end{eqnarray}

(2) Bob sends $N$ qubits $C_1\ldots C_N$ to Alice.

(3) After receiving the ancilla qubits $C_1\ldots C_N$, Alice
performs $U(f,G)$ on qubits $C_1\ldots C_NA_1\ldots A_M$. The state
of the composite system becomes
\begin{eqnarray}
\sum_{j=0}^{2^N-1}\alpha_j |j\rangle_{B_1\ldots B_N}
|f(j)\rangle_{C_1\ldots C_N} G(j)|\xi_j\rangle_{A_1\ldots A_M}.
\end{eqnarray}

(4) Alice performs a Hadamard transform on each qubit $C_i$, and
then measures $C_i$ in the computational basis. After the Hadamard
transform, the state of the composite system becomes
\begin{eqnarray}
\frac{1}{\sqrt{2^N}}\sum_{k=0}^{2^N-1} |k\rangle_{C_1\ldots C_N}
\sum_{j=0}^{2^N-1} (-1)^{f(j)\cdot k } \alpha_j |j\rangle_{B_1\ldots
B_N} G(j)|\xi_j\rangle_{A_1\ldots A_M},
\end{eqnarray}
where $f(j)\cdot k$ means the inner product modulo 2 of bit vectors
$f(j)$ and $k$.

If the measurement result of qubits $C_1\ldots C_N$ is
$k=k_1...k_N$, the state of $B_1\ldots B_NA_1\ldots A_M$ becomes
\begin{eqnarray}
\sum_{j=0}^{2^N-1} (-1)^{f(j)\cdot k } \alpha_j |j\rangle_{B_1\ldots
B_N} G(j)|\xi_j\rangle_{A_1\ldots A_M}.
\end{eqnarray}

(5) Alice informs Bob of the measurement result $k$ by sending $N$
cbits.

(6) Since Bob knows which restricted set $f$ the operation $U(f,G)$
belongs to, he can construct a corresponding $N$-qubit unitary
operation \begin{eqnarray}V(f)= \sum_{x=0}^{2^N-1}
|f(x)\rangle\langle x|.\end{eqnarray} Bob first performs $V(f)$ on
qubits $B_1\ldots B_N$, and then performs $\sigma_z =
|0\rangle\langle 0| - |1\rangle\langle 1|$ on qubit $B_i$ if and
only if $k_i=1$. The state of qubits $B_1\ldots B_NA_1\ldots A_M$
becomes
\begin{eqnarray}
\sum_{j=0}^{2^N-1} \alpha_j |f(j)\rangle_{B_1\ldots B_N}
G(j)|\xi_j\rangle_{A_1\ldots A_M} = U(f,G)
|\varphi\rangle_{B_1\ldots B_NA_1\ldots A_M}.
\end{eqnarray}

Thus, $U(f,G)$ has been successfully implemented on qubits
$B_1\ldots B_NA_1\ldots A_M$. The required communication resources
are $N$ qubits transmitted from Bob to Alice in step 2 and $N$ cbits
transmitted from Alice to Bob in step 5. The protocol in
Ref.~\cite{ZW08} which implements the same operations requires $N$
shared ebits and communication of $N$ cbits in each direction.

In the second case, if all $N+M$ qubits $B_1\ldots B_NA_1\ldots A_M$
are initially possessed by Bob, he has to send qubits $A_1\ldots
A_M$ together with qubits $C_1\ldots C_N$ to Alice in step 2. After
performing $U(f,G)$ on qubits $C_1\ldots C_NA_1\ldots A_M$, Alice
has to send qubits $A_1\ldots A_M$ back to Bob in step 5. In this
case, the required communication resources are $N+M$ qubits
transmitted from Bob to Alice plus $M$ qubits transmitted from Alice
to Bob plus $N$ cbits transmitted from Alice to Bob. The protocol in
Ref.~\cite{ZW07} which deals with this case requires $N+2M$ shared
ebits and communication of $N+2M$ cbits in each direction.

In the third case, if $M=0$, the restricted sets are reduced to
$U(f,\phi)$. In this case, Alice implements an $N$-qubit operation
on $N$ qubits possessed by Bob. The required communication resources
are $N$ qubits transmitted from Bob to Alice and $N$ cbits
transmitted from Alice to Bob. The protocol in Ref.~\cite{W06} which
deals with this case requires $N$ shared ebits and communication of
$N$ cbits in each direction.

\section{Communication Complexity Analysis}
\label{analysis}

In this section, we go on to discuss the communication complexity of
implementing restricted sets of multiqubit operations on distant
qubits. In the theory of quantum communication complexity, two
communication scenarios are often compared. In the
\emph{qubit-transmission scenario}, introduced by Yao \cite{Y93},
the parties can communicate qubits but are not allowed to share
prior entanglement in the initialization phase. In the
\emph{shared-entanglement scenario}, introduced by Cleve and Buhrman
\cite{CB97}, the parties have an initial supply of shared
entanglement but they can only communicate classical bits. In this
paper, we refer to a protocol as a \emph{qubit-transmission
protocol} if it requires (and only requires) transmission of qubits
and cbits, and refer to a protocol as a \emph{shared-entanglement
protocol} if it requires (and only requires) prior sharing of
entanglement and transmission of cbits.

The protocols in Refs.~\cite{EJPP00,W06,ZW07,ZW08} are
shared-entanglement protocols, and the required communication resources
of these protocols are ebits and cbits, whereas the protocol in
Ref. ~\cite{Y08} and our protocols are qubit-transmission protocols, and
the required communication resources of these protocols are qubits
and cbits. The required communication resources of these two kinds
of protocols can not be compared directly. They can only be compared
in the same communication scenario. Any qubit-transmission protocol
that requires the transmission of $N$ qubits and $M$ cbits can be
simulated in the shared-entanglement scenario through quantum
teleportation \cite{BBCJPW93} at the cost of $N$ shared ebits and
transmission of $2N+M$ cbits. On the other hand, any
shared-entanglement protocol that requires $N$ shared ebits and
communication of $M$ cbits can be implemented in the
qubit-transmission scenario at the cost of communication of $N$
qubits and $M$ cbits, because one party can prepare a pair of
entangled qubits and then transmit one of them to distribute 1
shared ebit. Tables~\ref{tab1},~\ref{tab2},~\ref{tab3}
and~\ref{tab4} summarize the required communication resources of
these protocols in both scenarios. The term gap is defined as the
communication resources of the upper protocol minus that of the
lower protocol.

Table~\ref{tab1} shows the required communication resources of
implementing $CU(N-1,1)$ which performs an arbitrary unitary
operation on a target qubit possessed by Alice with distant $N-1$
control qubits possessed by Bob. In the qubit-transmission scenario,
the protocol in Ref.~\cite{Y08} can save $1$ cbit of communication
compared to the protocol in Ref.~\cite{EJPP00}. In the
shared-entanglement scenario, the simulation of the protocol in
Ref.~\cite{Y08} requires $1$ more cbit of communication compared to
the protocol in Ref.~\cite{EJPP00}.

Table~\ref{tab2} shows the required communication resources of
implementing Alice's $(N+M)$-qubit operation on Bob's $N$ qubits and
Alice's $M$ qubits. Our protocol can save $N$ cbits of communication
compared to the protocol in Ref.~\cite{ZW08} in the
qubit-transmission scenario. Table~\ref{tab3} shows the required
communication resources of implementing Alice's $(N+M)$-qubit
operation on Bob's $N+M$ qubits. Our protocol can save $N+4M$ cbits
of communication compared to the protocol in Ref.~\cite{ZW07} in the
qubit-transmission scenario. Table~\ref{tab4} shows the required
communication resources of implementing Alice's $N$-qubit operation
on Bob's $N$ qubits. Our protocol can save $N$ cbits of
communication compared to the protocol in Ref.~\cite{W06} in the
qubit-transmission scenario. In summary, our protocols requires less
communication resources than the protocols in
Ref.~\cite{W06,ZW07,ZW08} implemented in the qubit-transmission
scenario.

In the shared-entanglement scenario, the right parts of
tables~\ref{tab2},~\ref{tab3} and~\ref{tab4} show that the
simulation of our protocols requires $N$ more cbits of communication
compared to the protocols in Ref.~\cite{W06,ZW07,ZW08}.

Therefore the saving of communicated cbits in our protocols is at
least $N$ cbits and goes up to $N+4M$ cbits when Alice's
$(N+M)$-qubit operation needs to be implemented on Bob's $N+M$
qubits. More essentially, our protocol has advantage because it is a
lot easier to transmit qubits than distribute and store entanglement
pairs.

Alice's multiqubit operations on Bob's qubits can be implemented
without prior sharing of entanglement as follows: Bob sends his
qubits to Alice, and then Alice applies the operation and sends them
back to Bob. By using this simple method, no auxiliary qubits are
used, no additional CNOT and Hadamard operations are required, and
no classical communications between Alice and Bob are needed.
However, our method requires fewer qubits to be communicated than
this simple method. This trade-off between computation and
communication in our method is analogous to the trade-off between
time and space in the field of algorithm design. Different benefits
are required in different situations. Sometimes we need an
easy-to-implement protocol, and sometimes we need to communicate as
few qubits as possible. The shared-entanglement protocol in
Ref.~\cite{ZW08} is fit for the parties who already have shared
entanglement. The simple method is fit for the case where additional
quantum operations are undesirable. However, our method is fit for the case where we place a high price on communication.

\begin{table}
\caption{$N$-qubit controlled-$U$ operation with Bob's $N-1$ control
qubits and Alice's 1 target qubit}
\label{tab1}       % Give a unique label
\begin{tabular}{c|cc|cc}
\hline
& \multicolumn{2}{|c|}{Qubit-transmission scenario}  & \multicolumn{2}{|c}{Shared-entanglement scenario} \\
& qubits & cbits & ebits & cbits \\
\hline
Protocol in Ref.~\cite{EJPP00} & $1$ & $2$ & $1$ & $2$ \\
Protocol in Ref.~\cite{Y08} & $1$ & $1$ & $1$ & $3$ \\
\hline Gap & & $1$ &  &
$-1$\\
\hline
\end{tabular}
\end{table}

\begin{table}
\caption{Alice's $(N+M)$-qubit operation on Bob's $N$ qubits and
Alice's $M$ qubits}
\label{tab2}       % Give a unique label
\begin{tabular}{c|cc|cc}
\hline
& \multicolumn{2}{|c|}{Qubit-transmission scenario}  & \multicolumn{2}{|c}{Shared-entanglement scenario} \\
& qubits & cbits & ebits & cbits \\
\hline
Protocol in Ref.~\cite{ZW08} & $N$ & $2N$ & $N$ & $2N$ \\
Our protocol & $N$ & $N$ & $N$ & $3N$ \\
\hline Gap & & $N$ &  & $-N$\\
\hline
\end{tabular}
\end{table}

\begin{table}
\caption{Alice's $(N+M)$-qubit operation on Bob's $N+M$ qubits}
\label{tab3}       % Give a unique label
\begin{tabular}{c|cc|cc}
\hline
& \multicolumn{2}{|c|}{Qubit-transmission scenario}  & \multicolumn{2}{|c}{Shared-entanglement scenario} \\
& qubits & cbits & ebits & cbits \\
\hline
Protocol in Ref.~\cite{ZW07} & $N+2M$ & $2N+4M$ & $N+2M$ & $2N+4M$ \\
Our protocol & $N+2M$ & $N$ & $N+2M$ & $3N+4M$ \\
\hline Gap & & $N+4M$ & &
$-N$\\
\hline
\end{tabular}
\end{table}

\begin{table}
\caption{Alice's $N$-qubit operation on Bob's $N$ qubits}
\label{tab4}       % Give a unique label
\begin{tabular}{c|cc|cc}
\hline
& \multicolumn{2}{|c|}{Qubit-transmission scenario}  & \multicolumn{2}{|c}{Shared-entanglement scenario} \\
& qubits & cbits & ebits & cbits \\
\hline
Protocol in Ref.~\cite{W06} & $N$ & $2N$ & $N$ & $2N$ \\
Our protocol & $N$ & $N$ & $N$ & $3N$ \\
\hline Gap & & $N$ &  &
$-N$\\
\hline
\end{tabular}
\end{table}

\section{Generalization to $d$-dimensional operations}
\label{generalization} In this section, we generalize the protocol
proposed in section 2 to $d$-dimensional operations. The
$(N+M)$-qu$d$it quantum operation $U_d(f,G)$ can be expressed as:
\begin{eqnarray}
U_d(f,G)= \sum_{x=0}^{d^N-1} |f(x)\rangle\langle x| \otimes
G(x),\end{eqnarray}where $G(x)$'s are arbitrary $d^M\times d^M$
unitary matrices.

Suppose that Alice wants to implement $U_d(f,G)$ on $N$ qu$d$its
possessed by Bob and $M$ qu$d$its possessed by herself. The initial
state of the $N+M$ qu$d$its can be written as:
\begin{eqnarray}
|\varphi\rangle_{B_1\ldots B_NA_1\ldots A_M} =
\sum_{j=0}^{d^N-1}\alpha_j |j\rangle_{B_1\ldots
B_N}|\xi_j\rangle_{A_1\ldots A_M},
\end{eqnarray}where subscripts $A_1\ldots A_M$ ($B_1\ldots B_N$) denotes Alice's (Bob's) $M$ ($N$)
qu$d$its. The protocol consists of six steps:

(1) Bob introduces $N$ ancilla qu$d$its $C_1\ldots C_N$ initially
prepared in the state $|0\rangle^{\otimes N}$, and then performs $N$
generalized controlled-NOT (CNOT) gates \cite{ADGJ09}
\begin{eqnarray}|x\rangle_{B_j}|y\rangle_{A_j} \rightarrow
|x\rangle_{B_j}|x - y\ mod\ d\rangle_{A_j}\end{eqnarray} with $B_i$
as the control qu$d$it and $C_i$ the target qu$d$it. After this
step, the state of the composite system becomes
\begin{eqnarray}
\sum_{j=0}^{d^N-1}\alpha_j |j\rangle_{B_1\ldots B_N}
|j\rangle_{C_1\ldots C_N} |\xi_j\rangle_{A_1\ldots A_M}.
\end{eqnarray}

(2) Bob sends $N$ qu$d$its $C_1\ldots C_N$ to Alice.

(3) After receiving the ancilla qu$d$its $C_1\ldots C_N$, Alice
performs $U_d(f,G)$ on qu$d$its $C_1\ldots C_NA_1\ldots A_M$. The
state of the composite system becomes
\begin{eqnarray}
\sum_{j=0}^{d^N-1}\alpha_j |j\rangle_{B_1\ldots B_N}
|f(j)\rangle_{C_1\ldots C_N} G(j)|\xi_j\rangle_{A_1\ldots A_M}.
\end{eqnarray}

(4) Alice performs the one-qu$d$it quantum Fourier transform
\begin{eqnarray}
|x\rangle \rightarrow \frac{1}{\sqrt{d}} \sum_{y=0}^{d-1} e^{2\pi
ixy/d}|y\rangle\end{eqnarray} on each qu$d$it $C_i$, and then
measures $C_i$ in the computational basis. After the quantum Fourier
transform, the state of the composite system becomes
\begin{equation}
\frac{1}{\sqrt{d^N}}\sum_{k=0}^{d^N-1} |k\rangle_{C_1\ldots C_N}
\sum_{j=0}^{d^N-1} exp\Big[ \frac{2\pi
i}{d}\sum_{l=1}^{N}f(j)_lk_l\Big] \alpha_j |j\rangle_{B_1\ldots B_N}
G(j)|\xi_j\rangle_{A_1\ldots A_M}.
\end{equation}

If the measurement result of qu$d$its $C_1\ldots C_N$ is
$k=k_1...k_N$, the state of qu$d$its $B_1\ldots B_NA_1\ldots A_M$
becomes
\begin{eqnarray}
\sum_{j=0}^{d^N-1} exp\Big[ \frac{2\pi
i}{d}\sum_{l=1}^{N}f(j)_lk_l\Big] \alpha_j |j\rangle_{B_1\ldots B_N}
G(j)|\xi_j\rangle_{A_1\ldots A_M}.
\end{eqnarray}

(5) Alice informs Bob of the measurement result $k$ by sending
$\lceil N \log_2 d\rceil$ cbits.

(6) Since Bob knows which restricted set $f$ the operation
$U_d(f,G)$ belongs to, he can construct a corresponding $N$-qu$d$it
unitary operation \begin{eqnarray}V_d(f)= \sum_{x=0}^{d^N-1}
|f(x)\rangle\langle x|.\end{eqnarray} Bob first performs $V_d(f)$ on
qu$d$its $B_1\ldots B_N$, and then performs $S^{k_i}$ on qu$d$it
$B_i$, where
\begin{eqnarray}S = \sum_{x=0}^{d-1} e^{-2\pi ix/d}|x\rangle\langle
x|.\end{eqnarray} The state of qu$d$its $B_1\ldots B_NA_1\ldots A_M$
becomes
\begin{eqnarray}
\sum_{j=0}^{d^N-1} \alpha_j |f(j)\rangle_{B_1\ldots B_N}
G(j)|\xi_j\rangle_{A_1\ldots A_M} = U_d(f,G)
|\varphi\rangle_{B_1\ldots B_NA_1\ldots A_M}.
\end{eqnarray}

Thus, $U_d(f,G)$ has been successfully implemented on qu$d$its
$B_1\ldots B_NA_1\ldots A_M$. The required communication resources
are $N$ qu$d$its transmitted from Bob to Alice in step 2 and $\lceil
N \log_2 d\rceil$ cbits transmitted from Alice to Bob in step 5.

\section{Conclusion}
\label{conclusion} We have considered the implementation of Alice's
multiqubit operation from the restricted sets \cite{ZW07,ZW08} on
distant qubits possessed by Bob from a communication complexity
perspective. The restricted sets are $2^N\times 2^N$ block matrices
with just one nonzero block in any column or any row, every block of
which is a $2^M\times 2^M$ unitary matrix. Protocols for
implementing these restricted sets of multiqubit operations on
distant qubits using prior sharing of entanglement have been
proposed in Ref.~\cite{ZW07,ZW08}. Inspired by Yang's work
\cite{Y08} for constructing a nonlocal $N$-qubit controlled-$U$ gate
without prior sharing of entanglement, we have proposed a protocol
to complete the same tasks as those in
Ref.~\cite{HPV02,W06,ZW07,ZW08} without prior sharing of
entanglement. We have shown that our qubit-transmission protocol
requires less communication resources than the previous
shared-entanglement protocols \cite{W06,ZW07,ZW08} in the
qubit-transmission scenario.  Because it is a lot easier to transmit
qubits than distribute and store entanglement pairs, our protocol
has advantage in the case that the parties have no prior sharing of
entanglement. Furthermore, we have generalized our protocol to
$d$-dimensional operations.

\begin{acknowledgements}
We are very grateful to Professor Brandt, Editor-in-Chief, and
 the anonymous reviewers for their invaluable comments
and detailed suggestions that helped to improve the quality of this
paper. This work is supported in
part by the National Natural Science Foundation (Nos. 60873055, 61073054), the Natural Science Foundation of Guangdong Province of China (No. 10251027501000004), the Fundamental Research Funds for the Central Universities (No. 10lgzd12),
the Specialized Research Fund for
the Doctoral Program of Higher Education of China (No. 20100171110042), the Program for New Century Excellent Talents in University (NCET) of China,
and the project of  SQIG at IT, funded by FCT and EU FEDER projects Quantlog POCI/MAT/55796/2004 and
QSec PTDC/EIA/67661/2006, IT Project QuantTel, NoE Euro-NF, and the SQIG LAP initiative.\\

\end{acknowledgements}



\end{document}